\documentstyle [manuscript,aps] {revtex}

\begin{document}

\draft

\title{Proton Decay from Excited States in Spherical Nuclei
\footnote{Work partially supported by the IN2P3-IFIN-HH Collaboration
grant no.95-17}}

\author{\c S. Mi\c sicu,}
\address{ National Institute for Nuclear Physics and Engineering,
Bucharest-M\u agurele, POB MG-6, Romania,}
\author{N. C\^ arjan and P. Talou}
\address{ Centre d'Etudes Nucl\' eaires de Bordeaux Gradignan, 
BP 120 - F 33175 Gradignan Cedex, \\
France}

\date{May 1998}

\maketitle

\begin{abstract}
Based on a single particle model which describes the time evolution of
the  wave function during tunneling across a one dimensional potential 
barrier we study the proton decay of $^{208}$Pb from excited states with 
non-vanishing angular momentum $\ell$. Several quantities of interest in this 
process like the decay rate $\lambda$, the period of oscillation $T_{osc}$, 
the transient time $t_{tr}$, the tunneling time $t_{tun}$ and the average 
value of the proton packet position $ r_{av} $ are computed and compared 
with the WKB results.
\end{abstract}


\section{\bf Introduction}

The recent interest in nuclei beyond the proton drip line, prompted 
the study of proton emission phenomenon from the ground state of such
nuclei \cite{Hof89}. Like in the case of $\alpha$ decay, this 
process consists in the tunneling of the proton across a potential 
barrier. However for this kind of radioactivity it is vital to add 
the centrifugal potential to the Coulomb barrier since the majority 
of emitted protons are likely to decay from states with $\ell\neq 0$.

In last years a series of theoretical investigations, based on the 
quasi-classical approximation  have been carried 
out in this field  \cite{Kad,BMP92,ASN97}. The proton half-lives
of observed heavy proton emitters were calculated and compared with 
the experimental ones and in some cases a good agreement was found. 

In the Gamow approach the decay is treated as a stationary 
process, the penetrability being given by the ratio of probabilities 
of finding the quantum particle on each side of the potential barrier. 
In this image the dynamical aspects are neglected. However  
quantities like the tunneling time are important in decay processes, 
in fission or in fusion reactions. 
In order to include the time evolution of a wave packet propagating in 
a classically forbidden region one need to solve the time dependent 
Schr\" odinger equation (TDSE). 
There have been some attempts to incorporate the time-dependency in the 
study of the proton decay \cite{FH83} by expanding the wave function 
into a term describing the bound-state of the prepared nucleus 
$\psi_0(A)$ and a set of exit channel wave functions $\chi$ orthogonal 
to it which decomposed into the intrinsic states of the daughter nucleus 
$\psi(A-1)$ and of the proton $\phi_p$.
However, imposing that the decay width is smaller than the resonance 
(quasi-stationary) energy ($\Gamma\ll E_0 $) 
the treatment reduce to the stationary Schr\" odinger equation  for 
$\psi_p(r,t=0)$ with a complex energy $E_0 - i\Gamma/2$. Afterwards the 
usual computational procedure is carried on \cite{Vr76}.

In this paper, based on a previous application of TDSE to the study of 
$\alpha$ decay \cite{SC94,SCS94}, we address the question of proton-decay
from orbitals with $\ell\neq 0$ for the spherical nucleus $^{208}$Pb.
This process is different from the ground-state proton emission
beyond the proton drip line. However the theoretical techniques that we
employ in this paper can be applied in both cases without major changes.
In our time-dependent approach the description of the decaying system 
is fully contained in the state vector. The time evolution of 
this vector enables us to determine the decay probability at any moment $t$. 
We are interested to study the influence of the potential on 
the tunneling quantities and to determine to what extent the errors of the 
WKB method depends on the energy of the quasi-stationary state and on the 
angular momentum.

\section{\bf Dynamical approach to quantum tunneling}

The interaction between the proton and the daughter nucleus $^{207}$Tl is
described by an average Woods-Saxon (WS) field which accounts for the nuclear
potential
\begin{equation}
V_{N}(r) = -\frac{V_0}{1+\exp(\frac{r-R_{0}^N}{a})}
\end{equation}
a Coulomb potential, which is approximated by the interaction between the 
point proton and the uniformly charged spherical core of charge $Z-1$
\begin{eqnarray}
V_{C}(r) & = & -\frac{(Z-1)e^2}{R_{0}^C}\left [ 1 + {1\over 2}  
\left (1 - \left (\frac{r}{R_{0}^C}\right )^2\right)\right ]  , 
r\le R_{0}^C   \\ \nonumber
& = & \frac{(Z-1)e^2}{r} , \hskip 4.65cm r\ge R_{0}^C  
\end{eqnarray}
and the centrifugal barrier
\begin{equation}
V_{cf}(r) = \frac{\hbar^2}{2\mu r^2}\ell(\ell+1)
\end{equation}
The WS interaction is defined by the nuclear radius 
${R_{0}}^{N}=r_0 A^{1/3}$, with $r_0$= 1.25 fm, the diffuseness 
$a$=0.7 fm, and the depth of the central potential $V_{0}$= 58 MeV
\cite{Ro68}.
Note that in other papers, the depth of the potential is not fixed, 
its value being adjusted to reproduce the experimental energy of the
quasi-bound state  \cite{BMP92,ASN97}. The Coulomb radius is given 
by $R_{0}^{C}=r_0 (A-1)^{1/3}$.

The initial wave function of the proton was chosen to correspond to a 
quasi-stationary state of the potential 
$V(r)=V_{N}(r)+V_{C}(r)+V_{cf}(r)$, with positive energy $E > 0$. 
The decay width of such a metastable state can be calculated using an 
analytic method developed by Gurwitz et al. \cite{GK87}. In  this 
approach it is assumed that the proton occupies a bound eigenstate in 
the potential well $V(r)+\varepsilon (r)$ which represents a slight 
modification of $V(r)$. Therefore we take 
\begin{equation} 
\psi_{p}(r,t=0) = \phi_{nl}^{(V+\varepsilon)}
\end{equation} 
where, $\phi_{n}^{(V+\varepsilon)}$ is an eigenstate of energy 
$E_{n\ell}^{(V+\varepsilon)}$ corresponding to the Hamiltonian
\begin{equation}
H_{0} = -\frac{\hbar^2}{2\mu}\Delta + V(r) +\varepsilon(r)
\end{equation}
where $\mu$ is the reduced mass of the daughter-proton system. 
The modification $\varepsilon(r)$ reads
\begin{eqnarray}
\varepsilon(r) & =& V(r_{max}) + (r-r_{max})\tan\theta - V(r) , 
\hskip 0.85cm r\ge r_{max}\\
\nonumber
& = & 0  , \hskip 6.75cm r\le r_{max}
\end{eqnarray}
where $\theta$ gives the slope of the potential barrier beyond the point 
$r_{max}$ at which $V(r)$ attain its maximum. 

In what follows we shall consider only the wave functions 
$\phi_{n\ell}^{(V+\varepsilon)}$ with the highest eigenvalue  
$E_{n\ell}^{(V+\varepsilon)}$ bellow the barrier $V_B = V(r_{max})$.

In Table I we list the heights of the barriers $(V_B)$, their 
locations $(r_{max})$ and the selected eigenvalues $(E_{n\ell})$ for a 
given value of the angular momentum $\ell$.

The next step consists in the resolution of the time dependent 
Schr\"odinger equation:
\begin{equation}
i\hbar\frac{\partial}{\partial t}\psi_p(r,t) = H(r)\psi_{p}(r,t)
\end{equation}
where
\begin{equation}
H(r) = -\frac{\hbar^2}{2\mu}\Delta + V(r)
\end{equation}
A numerical procedure based on the iterated leap-frog method, 
provides the solution of this equation \cite{Ser92}. 
The equidistant spatial grid used in this method is large enough 
such that no interference of the outgoing and ingoing wave
functions will take place during the time intervals employed in the 
calculations. 
Once we get the wave-function which describes the time evolution of the
proton packet through the potential barrier we are able to compute relevant 
quantities of the decay process.

The tunneling probability can be expressed as the probability of finding the 
proton beyond a certain point $r_B$ on the border which separates the zone
inside the barrier from the external one
\begin{equation}
P_{TD}(r_B,t) = \int_{r_B}^{\infty}|\psi_p(r,t)|^2 r^{2}dr
\end{equation} 
The decay rate is calculated according to the relation
\begin{equation}
\lambda_{TD}(r_B,t) = \frac{\dot P_{TD}}{1-P_{TD}}
\end{equation}

It is also interesting to calculate the average value of the proton 
wave packet position operator, $r_{av}$, inside the nucleus 
\begin{equation}
r_{av}(r_B,t) = 
\frac{\int_{0}^{r_{B}} r|\psi_p(r,t)|^2 r^2 dr}
                  {\int_{0}^{r_{B}} |\psi_p(r,t)|^2 r^2 dr} 
\end{equation}

\section{\bf Numerical results and comparison with WKB}

In Figure 1 a, b  we present the time evolution of the proton wave
function $\psi_p(r,t)$ for two angular momenta $l=0,5$ at four
different moments. We see the tendency of the wave function to decrease its 
amplitude in the interior of the barrier when time goes on. 
Eventually we observe that the fraction of the wave function which 
passed across the barrier behaves like a spreaded wave packet (at least 
on the spatial interval that we considered, i.e. up to 60 fm). 
This tendency is more pronounced for the wave function with $\ell$=5, 
because it faces a thiner barrier (see Table 2).
It is worthwhile to mention that although the wave function amplitude 
decreases constantly in time its shape does not change to much inside the 
barrier. 
 
In figures 2 a, b the time-dependent decay rate $\lambda_{TD}$ is
ploted for angular momentum $\ell$ = 0, 2, 5 and 8 . 
As has been pointed in an earlier work \cite{SC94} the decay rate
undergoes two regimes. In the first one, $\lambda_{TD}$ 
oscillates but increases on the average. The fact that at small times the 
decay rate is not constant as characteristic for exponential decay, but varies 
with time is typical for a quantum mechanical description \cite{GK84}. 
This fact contrasts to the usual classical image which portrays the 
radioactive system as an ensamble of nuclei decaying independently one of 
each other with a probability which does not depend on time.  
In the second regime $\lambda_{TD}$ performs small fluctuations around 
a constant value, that we call asymptotic value $\lambda_{TD}(\infty)$. 

The decay rates presented in figure 2 a correspond to $r_B=11.6$ fm, i.e.
the inferior limit of integration is choosed to lay between the two turning 
points as can be observed from Table 2.   
If $r_{B}$ is increased, the irregularities occuring in the first regime 
are smoothed out. This fact is pictured in Figure 2 b, where $r_B$ is
choosed to be 25 fm.

In Figure 3 we represented the behaviour of the wave packet's average 
position inside a potential region defined by $r_B = 25$ fm. As in the case 
of the decay rate we deal with two regimes. Whereas in the first regime 
$r_{av}$ increases up to a certain limit, in the second one it performs 
small-amplitude fluctuations around this limiting value as we expected 
since, according to a previous remark, the wave function does not change 
to much its shape.
The period of oscillations in this second regime is denoted by 
$T_{osc}$ and its value can be deduced simply by measuring the distance
between two maxima of $r_{av}$ (right column in Fig.3). 
As has been noted in previous papers \cite{SC94,SCS94}, the
quasi-stationary state tends to penetrate the potential barrier by performing 
these small-amplitude oscillations instead of simply crossing the barrier from 
one side to the other. For this reason it makes sense to associate the 
frequency of collisions in the formula of Gamow with $1/T_{osc}$ instead of 
$1/T_{cross}$, which is defined by the quasi-classical expression 
(see eq.(13) bellow).   

In the WKB approximation the decay rate $\lambda_{ST}$ is constant, its
value resulting from the product of the barrier penetrability $p$ and 
the collision frequency $\nu$
\begin{equation}
\lambda_{ST} =  \nu\cdot p
\end{equation}
where $\nu$ is given by the inverse of the classical period of motion
\begin{equation}
T_{cross} = \frac{2\mu}{\hbar}\int_{r_{t1}}^{r_{t2}}\frac{dr}{k(r)}
\end{equation} 
the wave number $k(r)$ reading
\begin{equation}
k(r)=\left [ \frac{2\mu}{\hbar^2}(Q-V(r))\right ]^{1/2}
\end{equation}
and the penetrability
\begin{equation}
p = \exp\left( -2\int_{r_{t2}}^{r_{t3}}d r
\sqrt{\frac{2\mu}{\hbar^2}(V(r)-Q)} \right )
\end{equation}


The stationary states computed in the modified
potential, become quasi-stationary when we turn on the real
potential (without the modification $\varepsilon (r)$) and their
energy is no longer well defined. Therefore, in all the above formulas
the decay energy $Q$ was taken to be the energy of the quasi-stationary
state $E_{0}$, computed as the average energy of the time dependent
Scr\" odinger equation
\begin{equation}
E_0=\langle \psi_p(r,t)| H(r) |\psi_p(r,t)\rangle
\end{equation}

In what concerns the decay rate, it can be inferred from Table 2 that 
the decay rates calculated in the time dependent approach $\lambda_{TD}$ 
are in relatively good agreement with the WKB values $\lambda_{ST}(\infty)$ 
especially when the quasi-stationary energy decreases with respect to the 
barrier height. Except the last case, with the highest angular momentum, in 
all other cases the WKB results overestimates the time-dependent values.

The comparison between the times $T_{osc}$ and $T_{cross}$ (see Table 2) 
shows that they have very close values. However they are refering to 
different types of motion, the first describing the small oscillations of 
the proton wave function between the walls of the barrier during  tunneling, 
the second, the classical movement inside the potential well. 

The transient time is defined as the time interval between the moment 
when $\lambda_{TD}$ starts to increase up to the moment when it reaches  
the limiting value $\lambda_{TD}(\infty)$. It depends on the energy of the  
quasi-stationary state. Its value can be deduced by inspecting 
figures 2 a and b.   
The tunneling time is related to the shift in time of the transition point 
between the two above mentioned regimes, for the decay rate. 
Therefore it can be associated to the time necessary
for the wave function to cross the barrier. Computing the decay rate for two 
different values of $r_B$, i.e. the barrier's turning points, and measuring 
the time delay between the two maxima of the two curves one gets the 
tunneling time (see Fig.4 a and b ). Notice that for $\ell=0,2,5$ the 
tunneling time decreases with increasing 
$\lambda_{TD}(\infty)$. However this does not happens for the state 
with higher angular momentum $\ell=8$, where, although the decay rate 
is smaller, the tunneling seems to take place faster.  
In fact the tunneling time is correlated with the imaginary time, which 
is nothing else than the time necessary for the proton packet to cross    
the inverse potential, i.e. 
\begin{equation}
t_{imag} = \int_{r_{t2}}^{r_{t3}}
dr~\sqrt{\frac{\mu}{2(V(r)-Q)}} 
\end{equation} 

\section{\bf Summary}

Motivated by recent theoretical and experimental investigations on 
proton radioactivity phenomenon, we studied the time dependent 
characteristics of the proton tunneling from excited states in the
spherical nucleus $^{208}$Pb. 
Although this process is different from the ground-state proton emission
beyond the proton drip line, the approach employed in this paper could be
easily extended to the above mentioned case of proton radioactivity.

Since other theoretical approaches are based on the 
semiclassical approximation we were interested to compare our 
exact results with the WKB ones.
We found that the discrepancy between the two methods decreases when 
the difference between the top of the barrier and the energy of the 
quasi-stationary state increases.
Our study does not concern a certain angular momentum state which 
could be measured in the decay reaction $^{208}$Pb$^{*}$$\rightarrow^{207}$Tl
$+ p$. Rather for a fixed set of Woods-Saxon parameters we investigated 
the dependence of the proton tunneling on time choosing one quasi-stationary 
state for every angular momentum. It seems that the accuracy of the WKB 
approximation increases with $\ell$. For a comparison with the experiment 
one should fit some of the WS parameters, e.g. the potential depth, in such 
a way to reproduce the observed energy. However the present comparison  
between the WKB and the time-dependent approaches gives an idea of the 
error involved in the stationary approach and provides a good starting point 
for future investigations of proton decay using TDSE which could
eventually answer to some questions related to this  phenomenon.

{}
\newpage

\begin{table} 
\caption{ 
The values of the barrier heights $V_B$, their locations $r_{max}$, 
the selected eigenvalues $E_{n\ell}$ and the wave functions number of 
nodes for different angular momenta $\ell$.}
\label{table:a}
\begin{tabular}{c c c c c c c}
Angular Momentum & Number of nodes & $r_{max}$(fm) & $V_{B}$(MeV) & 
$E_{n\ell}$(MeV) \\ \hline
$\ell=0$ &  4  &  10.62 & 10.256  & 7.78 \\
$\ell=1$ &  3  &  10.55  & 10.631  & -0.04 \\
$\ell=2$ &  3  &  10.45  & 11.392 & 7.23 \\
$\ell=3$ &  2  &  10.32  & 12.560 & -2.06 \\
$\ell=4$ &  2  &  10.15  & 14.163 & 5.29 \\
$\ell=5$ &  2  &  10.00  & 16.233 & 12.53 \\
$\ell=6$ &  1  &  ~9.82  & 18.800 & 0.64 \\
$\ell=7$ &  1  &  ~9.62  & 21.916 & 8.44 \\
$\ell=8$ &  1  &  ~9.45  & 25.607 & 16.58 \\
\end{tabular}
\label{table:1}
\end{table}

\begin{table} 
\caption{The quasi-stationary energies $E_0$, the turning points of the 
potential ($r_{t1}, r_{t2}, r_{t3}$), the asymptotic value of the decay 
rate $\lambda_{TD}(\infty)$ and its WKB correspondent $\lambda_{ST}$, 
the oscillation period $T_{osc}$ and the crossing time $T_{cross}$.}
\label{table:b}
\begin{tabular}{c c c c c c c c}
$E_0$ (MeV) & $r_{t1}$ (fm)& $r_{t2}$ (fm) & $r_{t3}$ (fm) & 
$\lambda_{TD}(\infty)$ (s$^{-1}$) & $\lambda_{ST}$ (s$^{-1}$) &
$T_{osc}$ (s) & $T_{cross}$ (s) \\ 
\hline
7.71  &  -  & 9.24 & 14.98 & 1.25$\times 10^{+20}$ & 1.74$\times 10^{+20}$
&2.75$\times 10^{-22}$ & 2.95$\times 
10^{-22}$\\
7.19  & 1.70 & 8.90 & 17.15  &   1.90$\times 10^{+19}$ & 2.33$\times
10^{+19}$ &2.60$\times 10^{-22}$ & 2.74$\times 
10^{-22}$\\
12.45 & 3.56 & 8.74 & 13.12 & 2.23$\times 10^{+20}$ & 2.59$\times
10^{+20}$ &2.24$\times 10^{-22}$ & 2.43$\times 
10^{-22}$\\
16.55 & 5.31 & 7.89 & 13.69 & 2.79$\times 10^{+19}$ & 2.52$\times 
10^{+19}$ &1.50$\times 10^{-22}$ & $1.79\times 
10^{-22}$\\ \end{tabular}
\label{table:2}
\end{table}

\vskip 2truecm
\centerline {\bf Figure  Captions }
\vskip 1truecm
 
 ${\bf Fig.~1.}$ Time evolution of the squared wave function 
$|\psi_{p}(r,t)|^{2}$
for angular momentum (a) $\ell=0$ and (b) $\ell=5$  at moments $t=$ 0,
1.5$\cdot$10$^{-21}$, 3$\cdot$10$^{-21}$ and 4.5$\cdot$10$^{-21}$ seconds. 
In the second case the barrier through which the proton undergoes
tunneling is thiner and therefore the probability to find it 
outside the barrier at a later time is larger. The wave function amplitude 
decreases in time and later on, the part of the wave function which
already tunneled manifests itself as a well-spreaded wave packet. 
\vskip 1.0truecm

${\bf Fig.~2.}$ The time dependent decay rates $\lambda_{TD}$ for the 
four selected quasi-stationary states of angular momentum 
$\ell = 0, 2, 5, 8$. In the eq.(10) we choosed (a) $r_B=11.6$ fm 
and (b) $r_B=25$ fm.
In all four cases we observe that after a certain time the decay
rate will fluctuate around an asymptotic value. In the second case these
fluctuations are smoother. 
\vskip 1.0truecm

${\bf Fig.~3.}$ The average value of the wave packet position
operator $r_{av}$ for three quasi-stationary states ($\ell= 0, 2, 5$).
On the left side the asymptotic behaviour of $r_{av}$ is observed.
whereas on the right side we foccused on the small amplitude oscillations
of the wave packet on its way to tunneling. 
\vskip 1.0truecm

${\bf Fig.~4.}$ The decay rates for (a) $\ell=0,2$ and (b) $\ell=5,8$ 
when $r_{B}$ is choosed to be the internal turning point (full line) 
and the external turning point (dotted line). The difference between the 
maxima of the two curves gives the tunneling time $t_{tun}$.

\vskip 1.0truecm

\vfill \eject

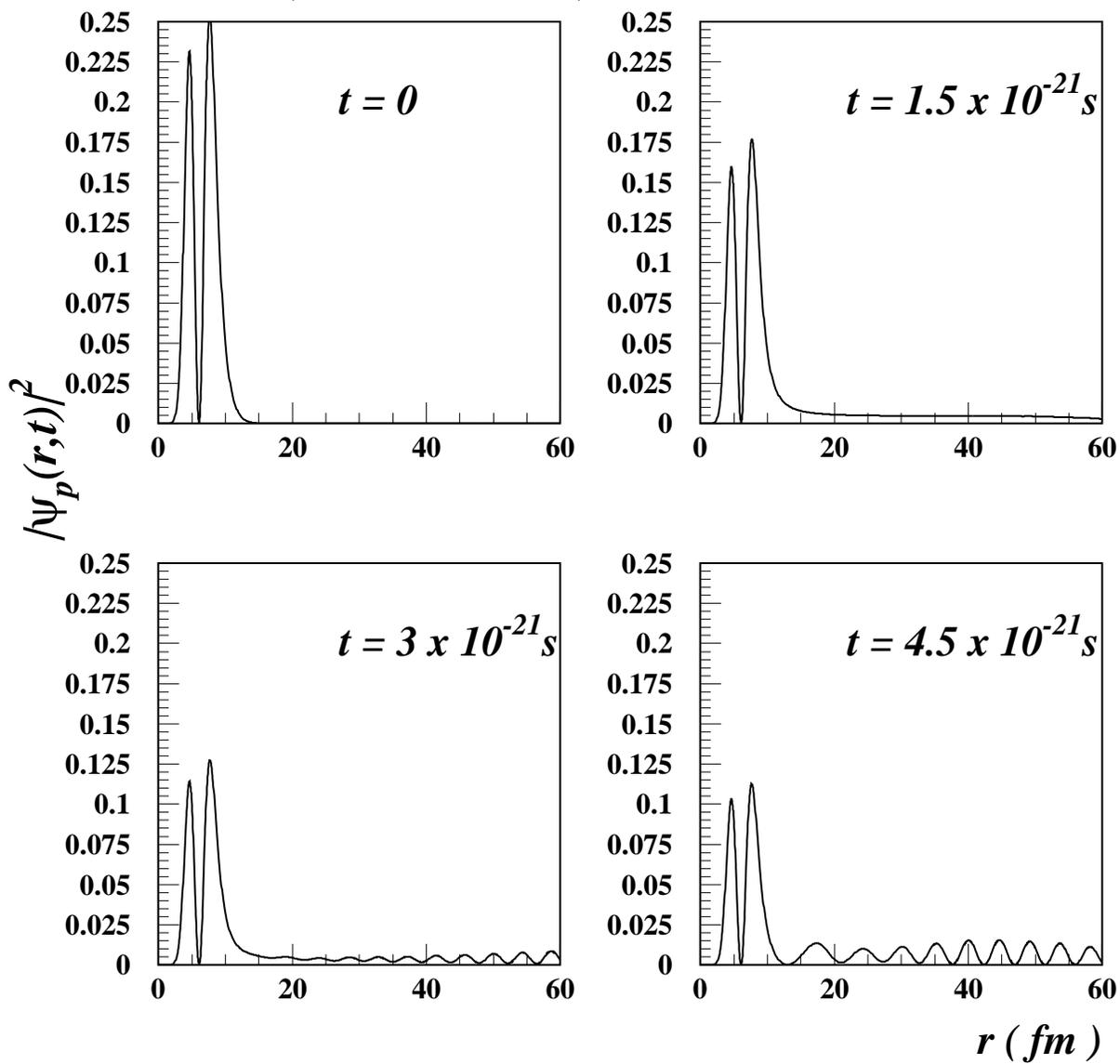
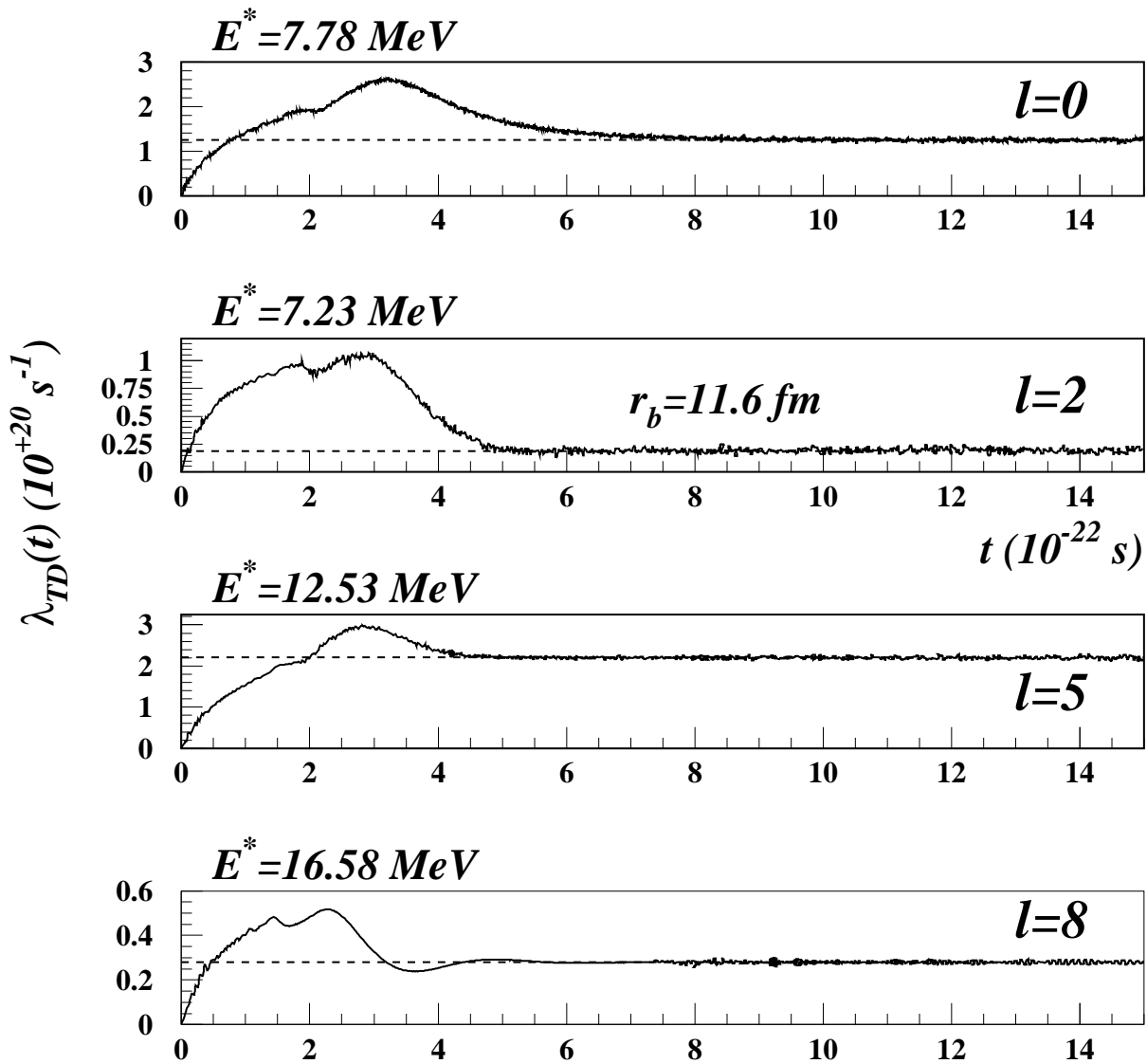
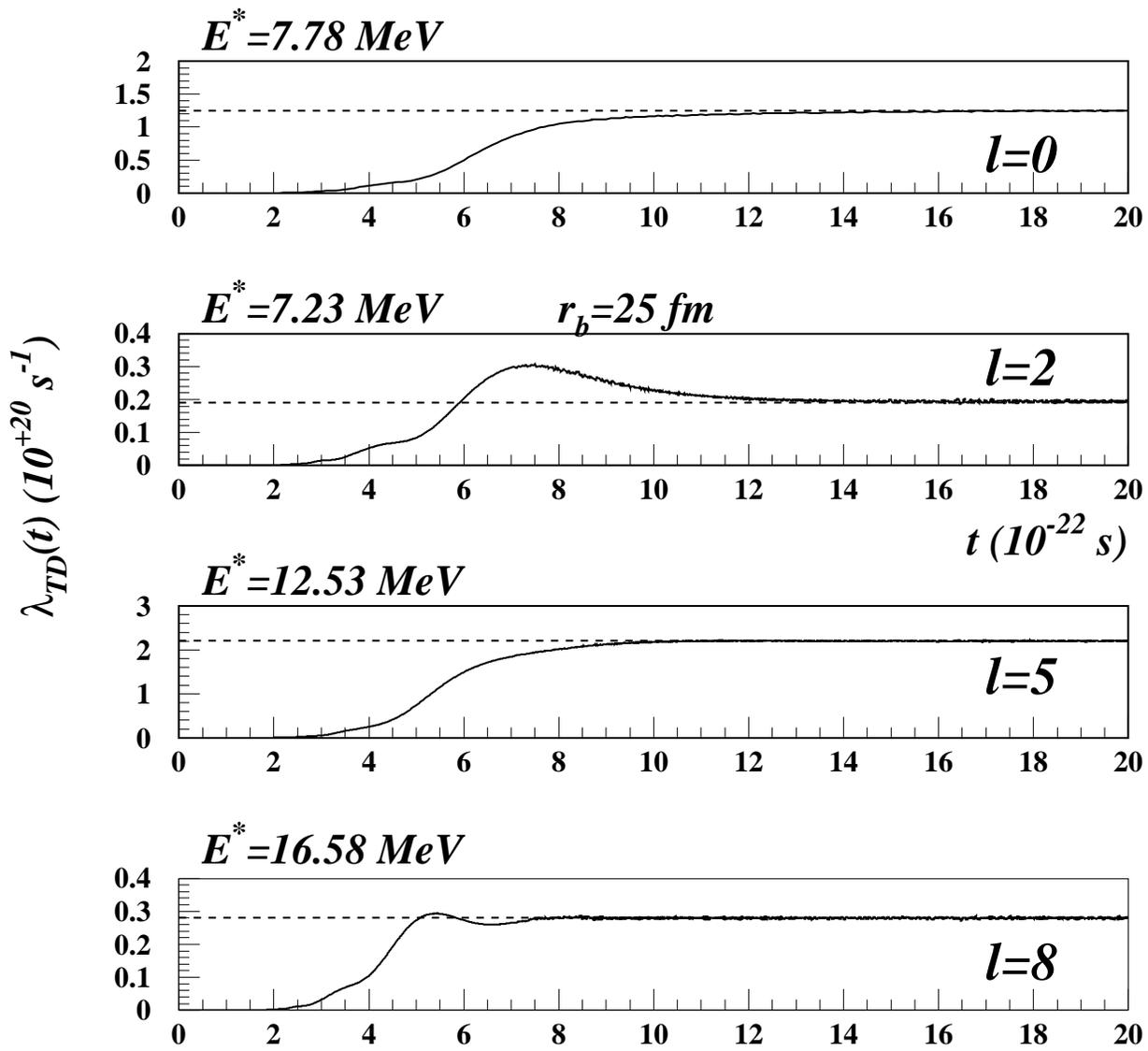
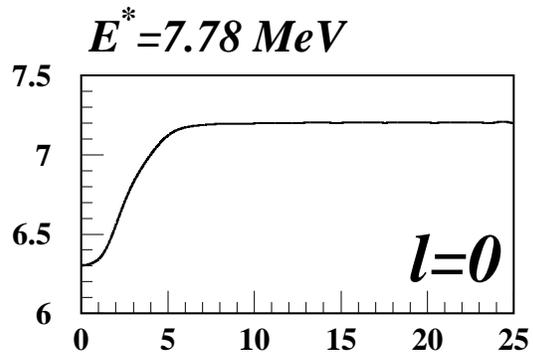
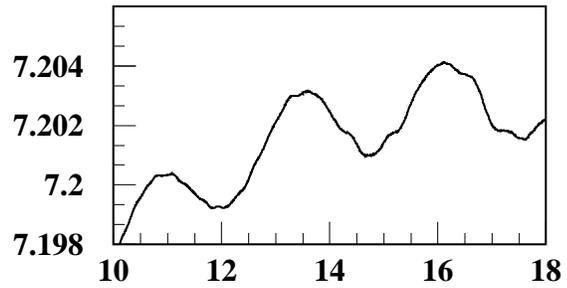
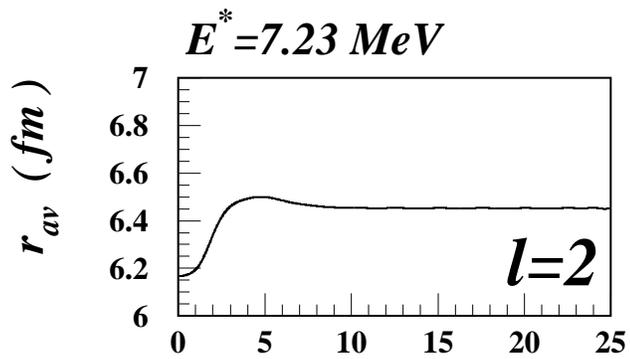
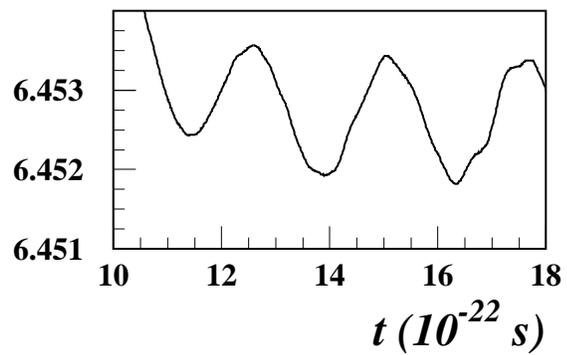
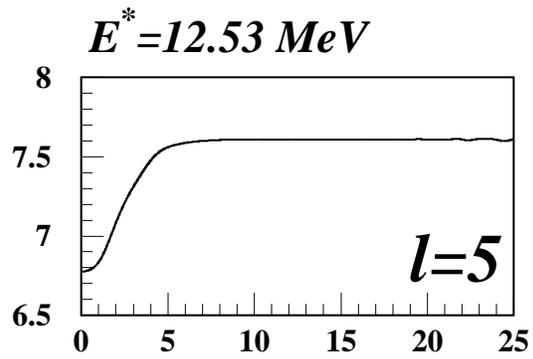
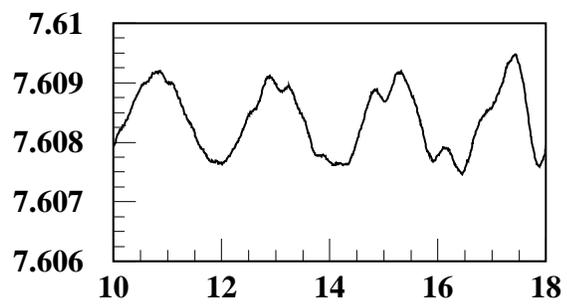
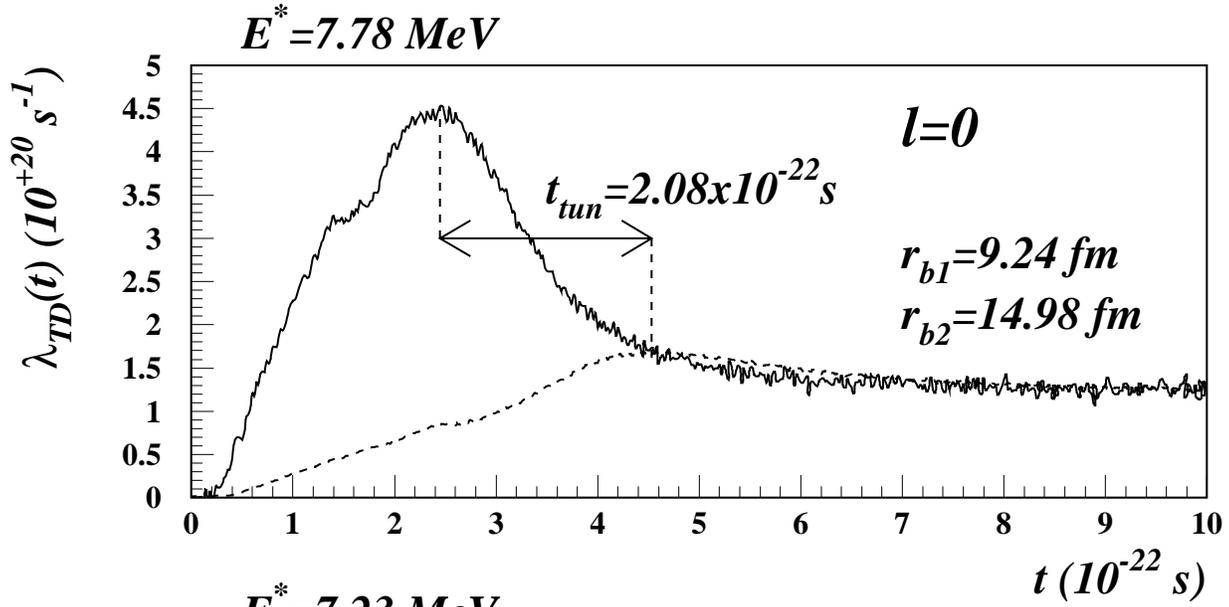
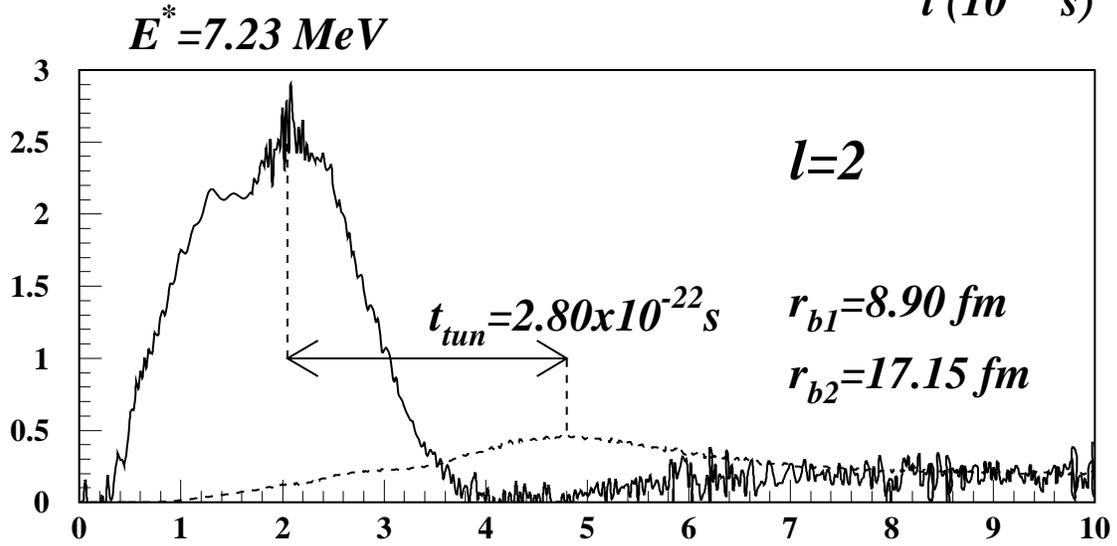
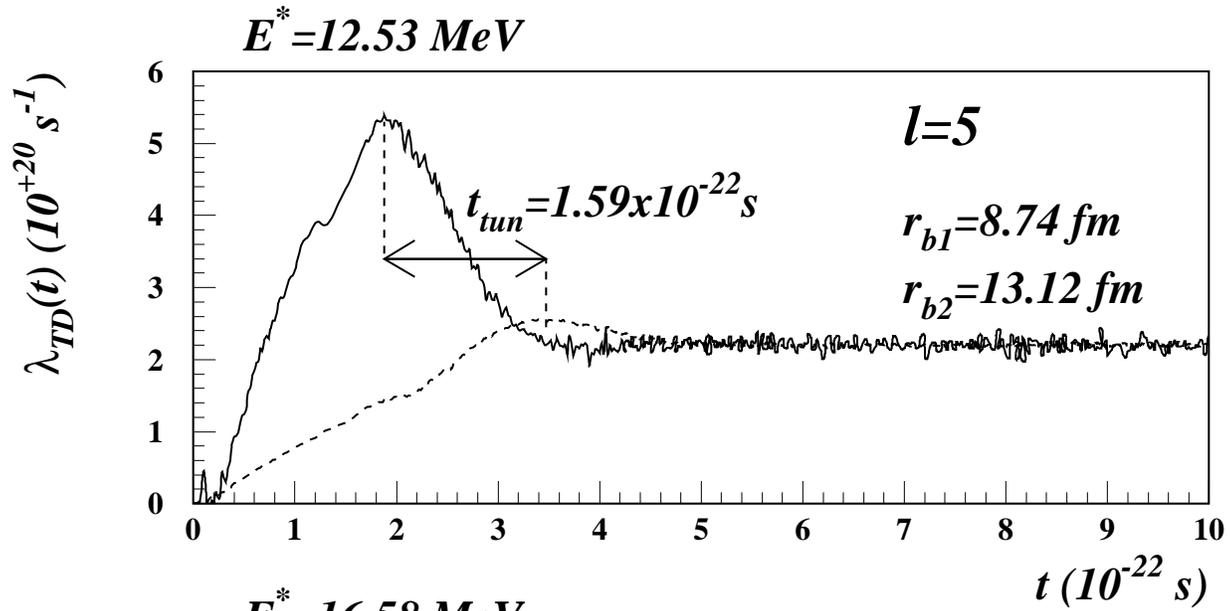
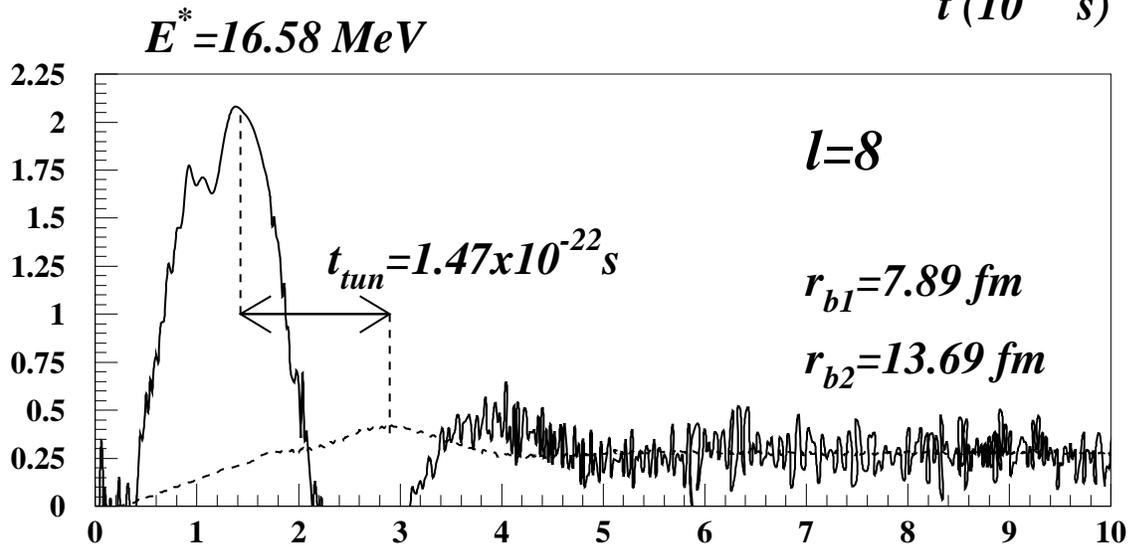
\end{document}